\documentclass[prb,aps,twocolumn,showpacs,superscriptaddress]{revtex4}
\usepackage{amsmath,amssymb}
\usepackage{graphicx}
\begin{document}

\title{Low-temperature magnetic properties of GdCoIn$_5$}

\author{D. Betancourth}
\affiliation{Centro At\'omico Bariloche (CNEA) and Instituto Balseiro (U. N. Cuyo), 8400 Bariloche, R\'io Negro, Argentina}

\author{J. I. Facio}
\affiliation{Centro At\'omico Bariloche (CNEA) and Instituto Balseiro (U. N. Cuyo), 8400 Bariloche, R\'io Negro, Argentina}

\author{P. Pedrazzini}
\affiliation{Centro At\'omico Bariloche (CNEA) and Instituto Balseiro (U. N. Cuyo), 8400 Bariloche, R\'io Negro, Argentina}

\author{C. B. R. Jesus}
\affiliation{Instituto de F\'isica ``Gleb Wataghin'', UNICAMP, Campinas-SP, 13083-859, Brazil}

\author{P. G. Pagliuso}
\affiliation{Instituto de F\'isica ``Gleb Wataghin'', UNICAMP, Campinas-SP, 13083-859, Brazil}

\author{V. Vildosola}
\affiliation{Departamento de Materia Condensada, GIyANN, CNEA, CONICET, 1650 San Mart\'in, Argentina}

\author{Pablo. S. Cornaglia}
\affiliation{Centro At\'omico Bariloche (CNEA) and Instituto Balseiro (U. N. Cuyo), 8400 Bariloche, R\'io Negro, Argentina}

\author{D. J. Garc\'ia}
\affiliation{Centro At\'omico Bariloche (CNEA) and Instituto Balseiro (U. N. Cuyo), 8400 Bariloche, R\'io Negro, Argentina}

\author{V. F. Correa}
\affiliation{Centro At\'omico Bariloche (CNEA) and Instituto Balseiro (U. N. Cuyo), 8400 Bariloche, R\'io Negro, Argentina}


\date{\today}


\begin{abstract}

A comprehensive experimental and theoretical study of the low temperature
properties of GdCoIn$_5$ was performed. Specific heat, thermal expansion, magnetization and electrical resistivity were measured in good quality single crystals down to $^4$He temperatures. All the experiments show a second-order-like phase transition at 30 K probably associated with the onset of antiferromagnetic order. 
The magnetic susceptibility shows a pronounced anisotropy below $T_N$ with an easy magnetic axis perpendicular to the crystallographic \^c-axis.
Total energy GGA+U calculations indicate a ground state with magnetic moments localized at the Gd ions and allowed a determination of the Gd-Gd magnetic interactions. Band structure calculations of the electron and phonon contributions to the specific heat together with Quantum Monte Carlo calculations of the magnetic contributions show a very good agreement with the experimental data.
Comparison between experiment and calculations suggests a significant anharmonic contribution to the specific heat at high temperature ($T \gtrsim$ 100 K).

\end{abstract}

\maketitle

\section{introduction}

GdCoIn$_5$ belongs to the family of rare earth (R) compounds RMIn$_5$ (M = Co, Rh, Ir). A vast set of low-temperature phenomenologies is observed in this family mostly associated with the 4$f$ electrons of R.\cite{Thompson2001,Pagliuso2006} This is particularly relevant for M = Co.
CeCoIn$_5$, for instance, is an ambient pressure unconventional superconductor ($T_c$ = 2.3 K) emerging from a heavy fermion state.\cite{Petrovic2001} Strong magnetic fluctuations, however, give raise to unique superconducting properties: a first-order-like transition at the upper critical field $H_{c2}$,\cite{Murphy2002,Tayama2002,Bianchi2002} an exotic spin-polarized superconducting phase at low temperature and high magnetic field,\cite{Radovan2003,Bianchi2003} and multiple vortex lattice phases with different non-hexagonal symmetries.\cite{Bianchi2008}

When substituting Ce by other elements along the rare earth series, the 4$f$ electrons rapidly lose their itinerant character and become localized originating robust magnetic ground states when R = Nd, Sm, Tb, Dy, Ho, Er and Tm.\cite{Isikawa2004,Thomas2005,Inada2006,Hudis2006,Huy2009} 
PrCoIn$_5$\cite{Hudis2006} and YbCoIn$_5$,\cite{Huy2009} on the other hand, remain paramagnetic in the whole temperature range while no information about EuCoIn$_5$ is available.
 
The intrinsic anisotropy associated with the tetragonal crystal structure deeply influentiates the magnetic interactions. Indeed, the R-ion magnetic moment tends to have an Ising-like character.
The dominant exchange couplings are antiferromagnetic. However, the actual antiferromagnetic wavevector and the ordered moment direction vary along the R-series and both are sensitive to the crystal electric field (CEF).

In this work we present a detailed experimental and theoretical study of the low temperature properties of another member of the family, GdCoIn$_5$. Gd$^{3+}$ ion has zero orbital angular momentum so the CEF effects are expected to be negligible. In this sense, this compound allows us to study the magnetic properties of a strong magnetic moment system without the influence of surrounding ions.            
Different experiments show a second order transition at $T_N$ = 30 K associated with an antiferromagnetic order.  This magnetic order is very robust against an applied magnetic field as high as 16 Tesla.
Below $T_N$, the magnetic susceptibility is anisotropic with a magnetic easy axis perpendicular to the crystallographic \^c-axis.  
Electronic and phonon band ab-initio calculations as well as an effective model calculation for the magnetic interactions reproduce quite well the observed low temperature specific heat. Comparison between experimental and calculated specific heat suggests that a significant anharmonic contribution exists at high temperature ($T \gtrsim$ 100 K).

\section{Experimental details}

GdCoIn$_5$ single crystals were grown by the self flux technique using sample growth facilities at the Centro At\'omico Bariloche. An initial mixture ($\sim$ 6 g) of high purity elements in the proportion Gd:Co:In = 1:1:15-20 is placed in an alumina crucible and then vacuum sealed in a quartz tube. Mixture homogeneity is achieved by initially warming the capsule up to 1200 $^\circ$C for 4 hours. It is then quenched to 750 $^\circ$C and slowly cooled down to 450 $^\circ$C over 10 days. At this temperature a centrifugue is used to remove the yet liquid excess-indium.

\begin{figure}[t]
\includegraphics[width=\columnwidth]{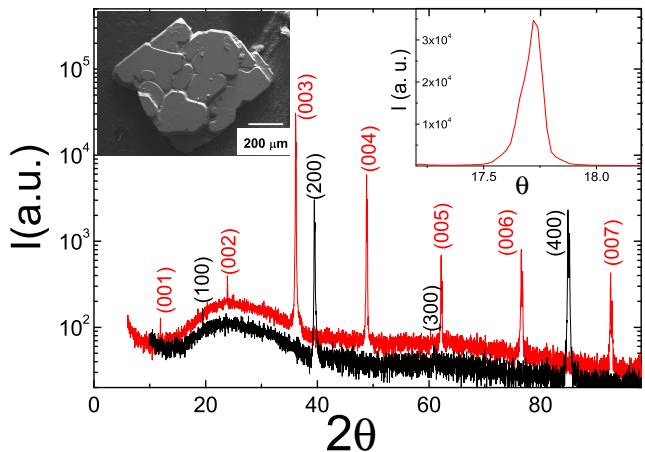}
\caption[]{(Color online) X-ray diffraction patterns for two sets of planes: ($h00$) and ($00l$). Insets: sample image taken with a SEM (left); rocking-curve for the (003) peak (right).}
\label{fig1}
\end{figure}

Platelet shaped single crystals of typical size 1$\times$1$\times$0.2-0.4 mm$^3$ were obtained (see inset to Fig. \ref{fig1}). Crystal structure and composition were examined by x-ray diffraction (XRD) and energy-dispersive x-ray spectroscopy (EDS) scans confirming the 1-1-5 stoichiometry. Figure \ref{fig1} shows typical XRD patterns corresponding to the ($h00$) and ($00l$) planes. The obtained lattice parameters are $a$ = 4.568(3) \AA$ $ and $c$ = 7.468(2) \AA. The rocking-curve FWHM for the (003) peak is as low as 0.11$^\circ$.

Magnetization experiments were performed in a Quantum Design MPMS. A high resolution ($\Delta L \leq$ 1 \AA) capacitive dilatometer was used in the thermal-expansion experiments. Specific heat was measured with both a bath modulation technique\cite{Graebner1989,Lortz2005} and a standard relaxation technique using a PPMS while a standard four-probe setup was used in the electrical transport measurements. 

\section{Results}

\begin{figure}[t]
\includegraphics[width=\columnwidth]{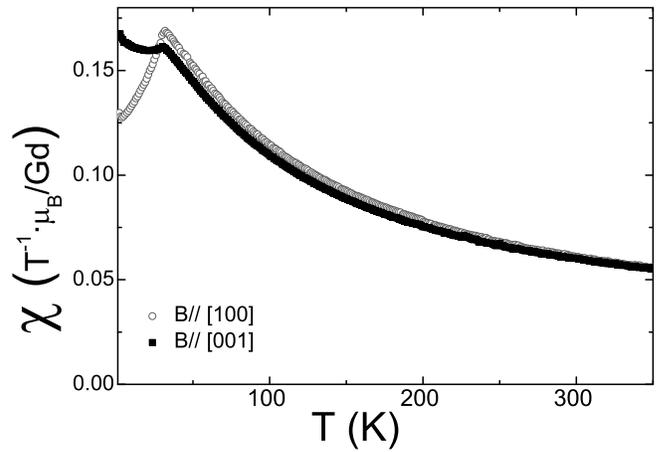}
\caption[]{(Color online) Temperature dependence of the experimental magnetic susceptibility along the [100] and [001] directions in an applied magnetic field $B =$ 1 T.}
\label{fig2}
\end{figure}

Figure \ref{fig2} shows the temperature dependence of the magnetic static susceptibility ($\chi = M/B$) along the [100] and [001] directions in an applied magnetic field $B =$ 1 T. A clear peak is observed along both axes at $T_N \approx$ 30 K. Remarkably, this ordering temperature is significantly lower than the $T_N$'s reported in the relatives GdRhIn$_5$ and GdIrIn$_5$, 40 K and 42 K, respectively.\cite{Pagliuso2001}
The high-$T$ susceptibility is mostly isotropic because the Gd$^{3+}$ ground state multiplet ($J$ = 7/2) remains almost unaffected by the crystal electric field since its orbital angular momemtum is zero. 
However, $\chi$ shows a clear anisotropy below $T_N$ indicating that the magnetic moments order parallel to the basal plane. This result is consistent with the magnetic structure determined in the relative compound GdRhIn$_5$.\cite{Granado2006}

The isotropic average high-$T$ susceptibility $\chi$ ($T >$ 150 K) can be fitted with the expression $\chi = \chi_{CW} + \beta$, where $\chi_{CW} = C$ / ($T - \theta$) is a Curie-Weiss-like susceptibility and $\beta =$ (0.020 $\pm$ 0.001) T$^{-1}\cdot$$\mu_B$/Gd.
An effective moment $\mu_{eff} =$ (7.9 $\pm$ 0.2)$\mu_B$/Gd and $\theta =$ (-46 $\pm$ 6) K are obtained.
The value of $\beta$ is much higher than any expected contribution from conduction electrons. It may be associated instead with an observed ferromagnetic background that is compatible with a 0.01 molar concentration of cobalt impurities.

\begin{figure}[h]
\includegraphics[width=\columnwidth]{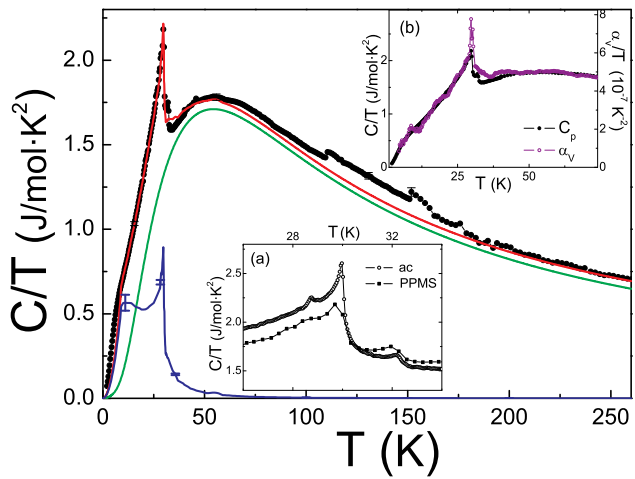}
\caption[]{(Color online) Temperature dependence of the experimental constant pressure specific heat, $C_p$ ($\blacksquare$) and the calculated phonon (green line), magnetic (blue line) and total (red line) specific heat including a correction for anharmonic effects. See text for details. Insets: (a) $C_p$ around the magnetic transition measured with two techniques (ac and relaxation); (b) $C_p$ and volume thermal-expansion $\alpha_v$ versus temperature.}
\label{fig3}
\end{figure}

Figure \ref{fig3} displays the temperature dependence of the specific heat at constant pressure, $C_p$, obtained with a PPMS. A clear second order transition is observed around 30 K (see inset (a) for $C_p$ close to the transition, measured with both the PPMS and a higher resolution bath modulation ac-technique) followed by a kink around 10 K, which is the typical behavior of a magnetic order arising from a high multiplet moment like Gd. There are two other small features at 28.5 K and 32 K whose origins are unknown at this point.

\begin{figure}[]
\includegraphics[width=\columnwidth]{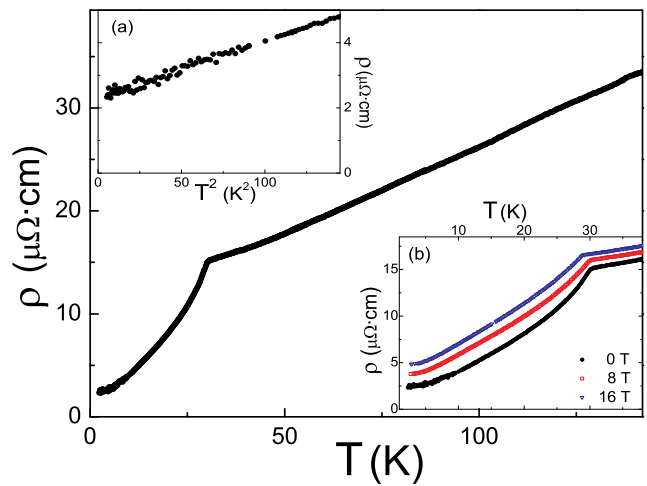}
\caption[]{(Color online) Temperature dependence of the in-plane electrical resistivity $\rho$. Inset (a): $\rho$ vs T$^2$. Inset (b): $\rho$ (T) at different applied magnetic fields.}
\label{fig4}
\end{figure}

The temperature dependence of the ab-plane electrical resistivity $\rho$ can be seen in Fig. \ref{fig4}. It has a linear dependence down to 50 K, below which there is a slight upturn down to $T_N$ = 30 K, most probably associated with magnetic fluctuations. At the transition temperature the resistivity has an abrupt decrease giving an extrapolated residual resistivity $\rho \left( 0 \right) \approx$ 2.2 $\mu \Omega\cdot$cm and a the residual resistivity ratio $RRR = \rho(300 K)/\rho(0) \approx$ 25. 
Below 12 K, $\rho \propto T^2$ as it is shown in the inset (a) to Fig. \ref{fig4}. 
The magnetic transition is very robust against an external applied magnetic field. Inset (b) to Fig. \ref{fig4} shows that $T_N$ is reduced by less than 2 K in a field of 16 Tesla perpendicular to the \^c-axis.

\section{Discussion}

An appropiate description of the low-temperature physical properties of GdCoIn$_5$ must consider at least three energy contributions: (i) the electronic configuration with a characteristic enery scale of several eV (the spread of the density of states); (ii) the lattice vibrations with an energy scale of some tenths of meV; and (iii) the magnetic interactions with an energy scale of about 3 meV (equivalent to the magnetic ordering temperature).
DFT calculations were performed within the generalized gradient approximation\cite{Perdew1996} with finite Coulomb repulsion $U$ for the Gd atoms (GGA+U).\cite{Facio2014} The method used is the full potential APW+local orbitals implemented in the Wien2k package\cite{Blaha1990} with the fully localized limit for the double counting correction. In relation to the Coulomb and exchange parameters, $U_{eff} = U - J$ = 6 eV was used as in bulk Gd.\cite{Yin2006,Petersen2006}

Magnetic properties of many 115 compounds are understood as resulting from the competition between three antiferromagnetic interactions: (i) in-plane nearest-neighbor ($K_0$), (ii) in-plane next-nearest-neighbor ($K_1$), and (iii) out-of-plane nearest-neighbor ($K_2$) couplings.\cite{vanHieu2006} All these are believed to be of the same order of magnitude causing a nearly frustrated magnetic system.
The results from GGA+U confirm this scenario were a three-couplings ($K_0$, $K_1$, $K_2$) model\cite{vanHieu2006} captures the main physics as the longer range couplings are an order of magnitude smaller. Moreover, the obtained couplings are distance-modulated strongly suggesting an RKKY origin.

The resulting magnetic model can be solved in the mean field approximation giving a second order phase transition to a C-type antiferromagnet at temperatures below $T_N^{MF} \approx$ 44 K. At the mean field level the only relevant parameter is an effective antiferromagnetic Gd-Gd exchange coupling $K_{MF} \sim$ 1.4 K, which determines the ground state energy and the transition temperature. In order to include quantum fluctuations in the calculations, a simplified model was considered consisting in an isotropic cubic lattice with nearest-neighbour Heisenberg exchange coupling $K_{eff}= K_{MF}$

\begin{equation}\label{eq:magmodel}
H_{m} = K_{eff} \sum_{\langle i,j\rangle} \vec{J}_i \cdot \vec{J}_j
\end{equation}

The simplifications to the original model allowed us to obtain numerical solutions through Quantum Monte Carlo (QMC) calculations. The QMC simulations were performed using the ALPS\cite{alps} library on a finite size lattice $L^3$ $(L = 12 - 30)$ and the results were extrapolated to $L \to \infty$.
The obtained QMC transition temperature $T_N^{QMC} \approx 32.3$ K is, as expected, lower than the mean field value. In order to calculate the magnetic contribution to the thermodynamic quantities and compare it with the experimental data, a value of $K_{eff} = 1.31$ K was used since it reproduces the experimental transition temperature.


From QMC, the magnetic contribution to the specific heat was calculated. This contribution is shown in Fig. \ref{fig3} (solid blue curve). An analytical contribution from antiferromagnetic magnons was added at low temperature ($T <$ 6 K).
The phonon spectrum was calculated using the Parlinksi-Li-Kawasoe method as implemented in the Phonopy code.\cite{Parlinski1997,Togo2008} The phonon density-of-states allowed us to determine the phonon contribution to the constant volume specific heat $C_v^h$ (green solid curve in Fig. \ref{fig3}).
This phonon contribution basically tracks the high temperature experimental features but it is shifted from the experimental curve by a temperature independent $C/T$ value. This value (55 mJ/mol$\cdot$K$^2$) is much higher than any expected contribution from conduction electrons ($\gamma =$ 6.2 mJ/mol$\cdot$K$^2$) as long as there is no evidence heavy fermion behavior.
 
Actually, that difference is indicative of a non-negligible anharmonic correction arising from the softening of the phonon frequencies. The main consequence of anharmonicity is that $C_p$ no longer coincides with $C_v$. In fact, $C_p$(300 K) = 186 J/mol$\cdot$K noticeably exceeds the Dulong-Petit value of 174.5 J/mol$\cdot$K.
On the assumption that at high temperatures the phonon frequencies $\omega_k$ explicitely depend on temperature as $\omega_k (T) = \omega_{k0} (1-\eta T)$, the anharmonic effects can be taken into account by a simple phenomenological correction $C_p=C^h_v (1+\eta T)$.\cite{Martin1991}
This correction is similar at high temperatures to the familiar expression $C_p - C_v = B_T V_m \alpha_v^2 \: \: T$, where $B_T$, $V_m$ and $\alpha_v$ are the isothermal bulk modulus, the molar volume and the volume thermal-expansion coefficient, respectively.
But in this last expression the constant volume specific heat $C_v$ is computed at the corresponding $V(T,P)$ including anharmonic terms while $C^h_v$ is obtained only from harmonic terms.\cite{Facio2014,Martin1991}

The resulting calculated total specific heat, including anharmonic effects and the magnetic contribution, is shown in Fig. \ref{fig3} (solid red curve). An excellent agreement with the experimental result is obtained when taking $\eta =$ 2.5 $\times$ 10$^{-4}$ K$^{-1}$. A similar value has been reported recently in a related compound.\cite{Cermak2013}
The large magnetic entropy at low temperature makes it difficult to extract a Debye temperature $\theta_D$ from the experimental $C_p$. 
Nonetheless, given the excellent agreement with the ab-initio results, we can use the calculated phonon contribution to obtain $\theta_D$ = 124.5 K. 
This value is in very good agreement with reported Debye temperatures in relative compounds of the same family.\cite{Hegger2000,vanHieu2007} 

The magnitude of the anharmonic effect is further confirmed by thermal expansion measurements. Inset (b) to Fig. \ref{fig3} shows the temperature dependence of $\alpha_v$, which basically tracks $C_p$ outside the magnetic transition. 
The GGA +U-calculated value for $B_T=$ 66.2 GPa\cite{Facio2014} (reported experimental value for the relative CeCoIn$_5$ is 72.8 GPa)\cite{Brady2013} gives a Gr\"uneisen parameter $\Gamma = \frac{V_m \alpha_v B_T}{C_v} \cong $ 2 at high-$T$ coinciding to what is expected for lattice vibrations.\cite{Moruzzi1988} 

\section{Conclusions}

A comprehensive study of the low temperature properties of GdCoIn$_5$ was performed. Good quality single crystals show an antiferromagnetic transition at $T_N =$ 30 K. 
This magnetic state is very robust against an external magnetic field.
Ab-initio and Quantum Monte Carlo calculations give a very good account of the experimental specific heat showing that anharmonic effects are important above 100 K.
The high-temperature magnetic susceptibility $\chi$ follows a Curie-Weiss law with $\mu_{eff} =$ (7.9 $\pm$ 0.2)$\mu_B$/Gd and $\theta =$ (-46 $\pm$ 6) K.  
Below $T_N$, $\chi$ shows a pronounced anisotropy with an easy magnetic axis along the basal plane. It is noteworthy to stress this observation. It suggests that, even though Gd$^{3+}$ ions have zero angular momentum, a tiny but non-negligible crystal electric field (CEF) effect may be present, and/or that a relevant magnetoelastic coupling could exist below $T_N$. 
Ongoing magnetostriction experiments together with a more sophisticated magnetic model including CEF effects and magnetostructural coupling could help to address this issue.

\section{Acknowledgments}

We aknowlegde B. Alascio for fruitful discussions.
Work partially supported by CONICET (through PIP0702, PIP0832, PIP00273), ANPCyT (through PICT07-00812), SeCTyP-UNCuyo (through 06/C393) and FAPESP (through TEMATICO 2021/04870-7). D. B. and J. I. F. hold scholarships from CONICET while P. P., V. V., P. S. C., D. J. G. and V. F. C. are members of CONICET, Argentina.

\end{document}